\documentclass[prb,aps,twocolumn,dcolumn,superscriptaddress,showpacs,amsfonts,amsmath,amssymb,showkeys,floatfix]{revtex4}

\usepackage{bm}
\usepackage{graphicx}

\usepackage{amssymb}

\begin{document}

\title{Phase transitions in systems of magnetic dipoles on a
square lattice with quenched disorder}
\date{\today}
\author{Juan J. Alonso}
\affiliation{F\'{\i}sica Aplicada I, Universidad de M\'alaga,
29071-M\'alaga, Spain}
\email[E-mail address: ] {jjalonso@uma.Es}

\date{\today}

\begin{abstract}
We study by Monte Carlo simulations the effect  of quenched orientational 
disorder in systems of interacting classical dipoles on a square lattice.
Each dipole can lie along any of two perpendicular axes that
form an angle $\psi$ with the principal axes of the lattice. 
We choose $\psi$ at random and without bias from the 
interval $[-\Delta, \Delta]$ for each site of the lattice. 
For $0 \le \Delta \lesssim \pi/4$ we find a 
thermally driven  second order transition  
between a paramagnetic and a dipolar antiferromagnetic  
order phase and critical exponents that change continously with $\Delta$. 
Near the case of maximum disorder $\Delta \approx \pi/4$ we still find 
a second order transition at a finite temperature $T_c$ but our results
point to {\it weak} instead of {\it strong} long-ranged dipolar order for 
temperatures below $T_c$. 
\end{abstract}

\maketitle

\section{Introduction}
Systems of interacting dipoles (SIDs) are atracting a renewed interest. 
This is in part due to recent advances in nanoscience \cite{nano} which
render available realizations of magnetic nanoparticle assemblies 
\cite{array}. These
systems show a rich collective behaviour in which the dipole-dipole
interactions play an essential role. 
 This is because dipole-dipole interaction strength grows 
linearly with nanoparticle volume. Collective effects that are 
controlled by dipolar interactions can therefore become 
important at reasonably high temperatures.
Spatial variations of the direction of magnetic dipolar fields 
lead to frustration and make SIDs very sensitive to their spatial 
arrangements \cite{our1}. In crystalline arrangements for example,  
SIDs exhibit different long ranged ferro or antiferromagnetic 
dipolar magnetic order that depends crucially on lattice geometry.  
The magnetic ordering depends also crucially
in anisotropy. On the one hand, single site anisotropy is always 
present in magnetic crystals. On the other hand, dipole-dipole 
interactions by themselves create effective anisotropies in 2D systems. 
In square lattices, for example, dipole-dipole interactions 
pushes spins to lie on the plane of  the lattice, and termal 
excitations tends to align them  along the two principal axes 
of  the lattice \cite{debell}. The resulting magnetic order
states that ensue from the competition of dipolar and anisotropic
energies could be as exotic as "spin ice" found in diamond type crystals
\cite{ice}.
SIDs in disordered spatial arrangements are also of interest. Some 
non equilibrium spin glass behaviour (like time dependent 
susceptibilities and aging) has been observed in experiments 
with systems of randomly placed dipoles 
as frozen ferrofluids or strongly diluted magnetic
crystals \cite{disorder}. Very recent computer 
simulations indicate that systems
of Ising dipoles with random anisotroy axes have an equilibrium 
spin glass phase at low temperatures\cite{jul}.

The aim of this paper is to study by numerical simulations
the effect of quenched directional disorder in a system 
of dipoles placed in cristalline array.
We are interested in how the magnetic dipolar order 
and the character of the transition between 
the paramagnetic and the low temperature ordered phase vary 
as we increase the amount of disorder. We also investigate 
whether there is a well defined threshold of 
disorder beyond which dipolar order dissapears.

\begin{figure}[!b]
\includegraphics*[width=88mm]{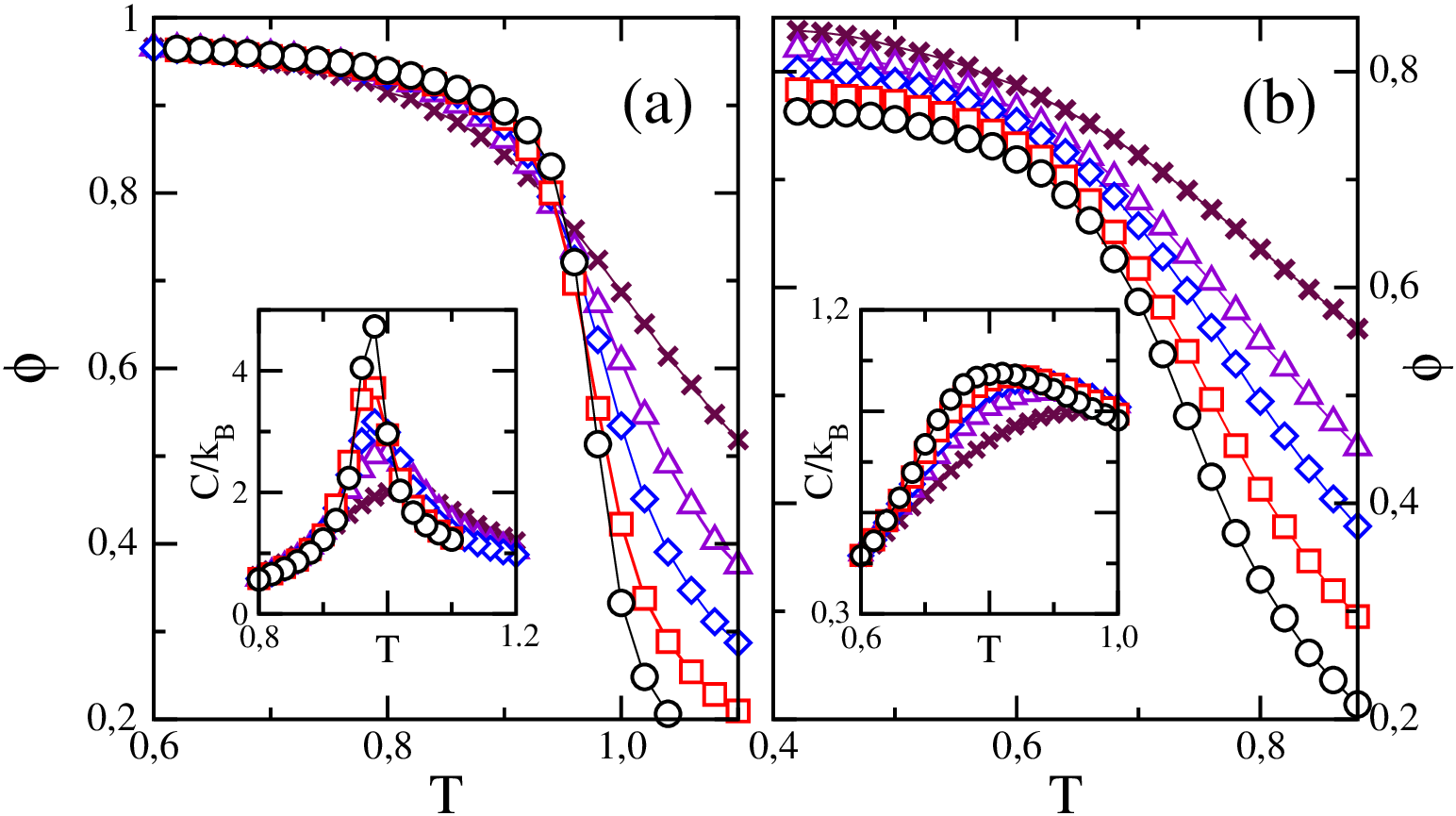}
\caption{(a) Order parameter vs $T$ for systems with 
$\Delta=0.4$.
$\times$, $\bigtriangleup$, $\diamond$, $\Box$ and  
$\circ$ 
stand for
$L=8,12,26,22$ and $32$ respectively. Lines are only guides to the eyes.
In the inset, specific heat vs. $T$ for the same $\Delta$ and systems
sizes. (b) and inset therein, same as in (a) but for $\Delta = \pi/4$.
} 
\label{fig1}
\end{figure}

\section{Model and calculation}
We first define the system model we use. Let ${\bf S}_i$ be a 
classical two xy-component unit spin at lattice site $i$ of a square 
lattice. These spins interact as pure magnetic dipoles with Hamiltonian 
\begin{equation}
H=\sum_{\langle ij\rangle}\sum_{\alpha\beta} 
T_{ij}^{\alpha\beta}S_i^\alpha S_j^\beta,
\label{H}
\end{equation}

\noindent where 

\begin{equation}
T_{ij}^{\alpha\beta}=\varepsilon_d
\left(\frac{a}{r_{ij}}\right)^3\left(\delta_{\alpha\beta}-3
\frac{r_{ij}^\alpha r_{ij}^\beta}{r_{ij}^2}\right),
\label{dipene}
\end{equation}
\noindent ${\bf r}_{ij}$ is the displacement from site $i$ to site $j$, $a$ is the square
lattice parameter. In the following all energies and temperatures are given in 
terms of $\varepsilon_d$ and $\varepsilon_d/k_B$ respectively.
When there is no quenched disorder, 
we consider a strong quadrupolar anisotropy that forces spins
to lie along any of the two principal axis of the square lattice. 
This dipolar four-state clock model has been studied recently, and 
found to have a second order transition between a paramagnetic and
an antiferromagnetic dipolar phase at $T_c=1.106$ with critical
exponents $\alpha/\nu=0.82(5)$ and $\beta/\nu=0.03(2)$ \cite{our2}.
In the ordered phase spins point up along lines with  alternate sign 
from one line to the adjacent one. In order to characterize these 
antiferromagnetic order, we define a staggered magnetization as 
\begin{equation}
(m_x,m_y)=N^{-1} (\sum_i S_i^x (-1)^{y(i)}, \sum_i S_i^y (-1)^{x(i)}),
\label{stag}
\end{equation}
and use $\phi=(m_x^2+m_y^2)^{1/2}$ as order parameter.

This system has been found to be very sensitive to the direction of 
the quadrupolar anisotropy axes. For example, if one turn these axes 
an angle $\psi =\pi/4$, critical exponents change to $\alpha/\nu=0.44(4)$ 
and $\beta/\nu=0.00(5)$\cite{unpub}. Thus, we 
introduce disorder in the model described above by tilting 
these two orthogonal anisotropy axes by an angle $\psi$ with respect
to the crystalline axes. For each spin this angle  $\psi$ is chosen 
at random and without bias from the interval 
$[-\Delta, \Delta]$.
Note that $\Delta=\pi/4$ correspond to a completely
random distribution of the orientation of these pairs of axes. Note also
that even in this case anisotropies never force any pair of spins to form angles
greater that $\pi/4$. Therefore, this is less disruptive than  
random {\it uniaxial} anisotropy \cite{jul}.

\begin{figure}[!b]
\includegraphics*[width=75mm]{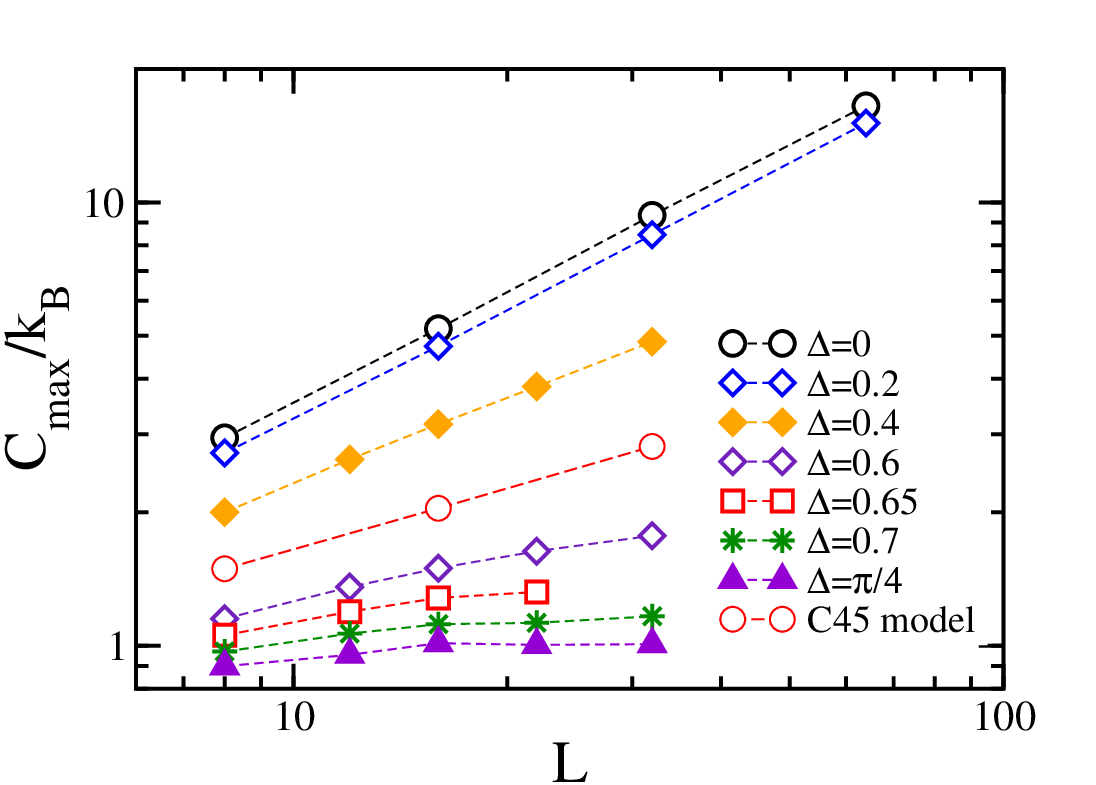}
\caption{ Log-log plot or the maximum value of the specific heat
as a function of $L$ for  the indicated values of $\Delta$.
$\bullet$ stands for the model without disorder, but
with quadrupolar easy axis forming angles of $pi/4$
with the $x,y$ axis at every site of the lattice.}
\label{fig2}
\end{figure}

We study systems of $L\times L$ spins and use periodic 
boundary conditions.  We let a spin on site $i$ interact 
through dipolar fields with all spins within
an $L\times L$ square centered on the $i$- site \cite{our2}.  
Our simulations follow the standard Metropolis Monte
Carlo (MC)  algorithm \cite{metrop}.
We use a single-spin flip dynamics, in which
all dipolar fields are updated throughout the system
every time a spin flip is accepted, before another spin
is chosen in order to repeat the process.
In our MC simulations, we 
start for a given realization of disorder
from $T=2.5$ (well in the paramagnetic phase) 
and lower temperature in $\Delta T=0.05$ steps. At each value of 
$T$ we averaged over $5 \times 10^6$ MC sweeps having first discarded
$10^6$ MC sweeps to let the system equilibrate. 
Finally, our results are averaged over $N_r$ different
independent realizations of quenched disorder. In order to get realiable
results we use at least $N_r=200$ for $\Delta > 0.6$ and $N_r=50$ for
$\Delta < 0.6$. As a result, error bars in the figures shown in this paper
are always smaller than symbol sizes used therein. Our results
follow from simulations for system sizes $L=8, 12, 16, 22$ and $32$. 
This would be insuficient to obtain accurate critical exponent values 
but is adequate for our purposes.

We study the modulus of the staggered magnetization $\phi$ defined 
above and  obtain the susceptibility from their fluctuations as 
$\chi =\delta \phi^2 /(NT)$. We calculate the specific heat from 
energy fluctuations via the relation $C=\delta E^2/(NT^2)$. Finally,
we use the cumulant like quantity

\begin{equation}
u_{12}=  [\pi/(4-\pi)][(4/\pi)-<\phi^2>/<\phi>^2]
\end{equation} 
As $L \to \infty$ we expect that $u_{12} \to 0$ in the
paramagnetic phase, $u_{12} \to 1$ in the long ranged
ordered phase and tend to some intermediate value at
a critical point. Hence, curves of $u_{12}$ versus $T$ for
different system sizes should cross at critical points for
large enough $L$. 

\section{Results}
Data obtained for the order parameter $\phi$ for $\Delta=0.4$
are exhibited in a $\phi$ versus $T$ plot in figure 1(a). 
Curves for different $L$
cross at $T_c=0.95(1)$. Below this temperature
$\phi$ increases as $L$ increases indicating the existence of
an ordered phase. Plots 
for $\Delta =0.2, 0.4, 0.65$ and $0.7$ (not shown) are
qualitalively similar, except in that curves for  
$L \ge 16$ tend to collapse below $T_c$ 
for the case $\Delta =0.7$. The behavior is markedly different for  
$\Delta \approx \pi/4$ (see Fig.~1(b)),
where $\phi$ appears to decrease as $L$ increases even at low $T$
at least for the sizes we have studied, raising the question over the existence of  
strong long-range order. We return to this point in the discussion
of Fig.~4.

\begin{figure}[!b]
\includegraphics*[width=88mm]{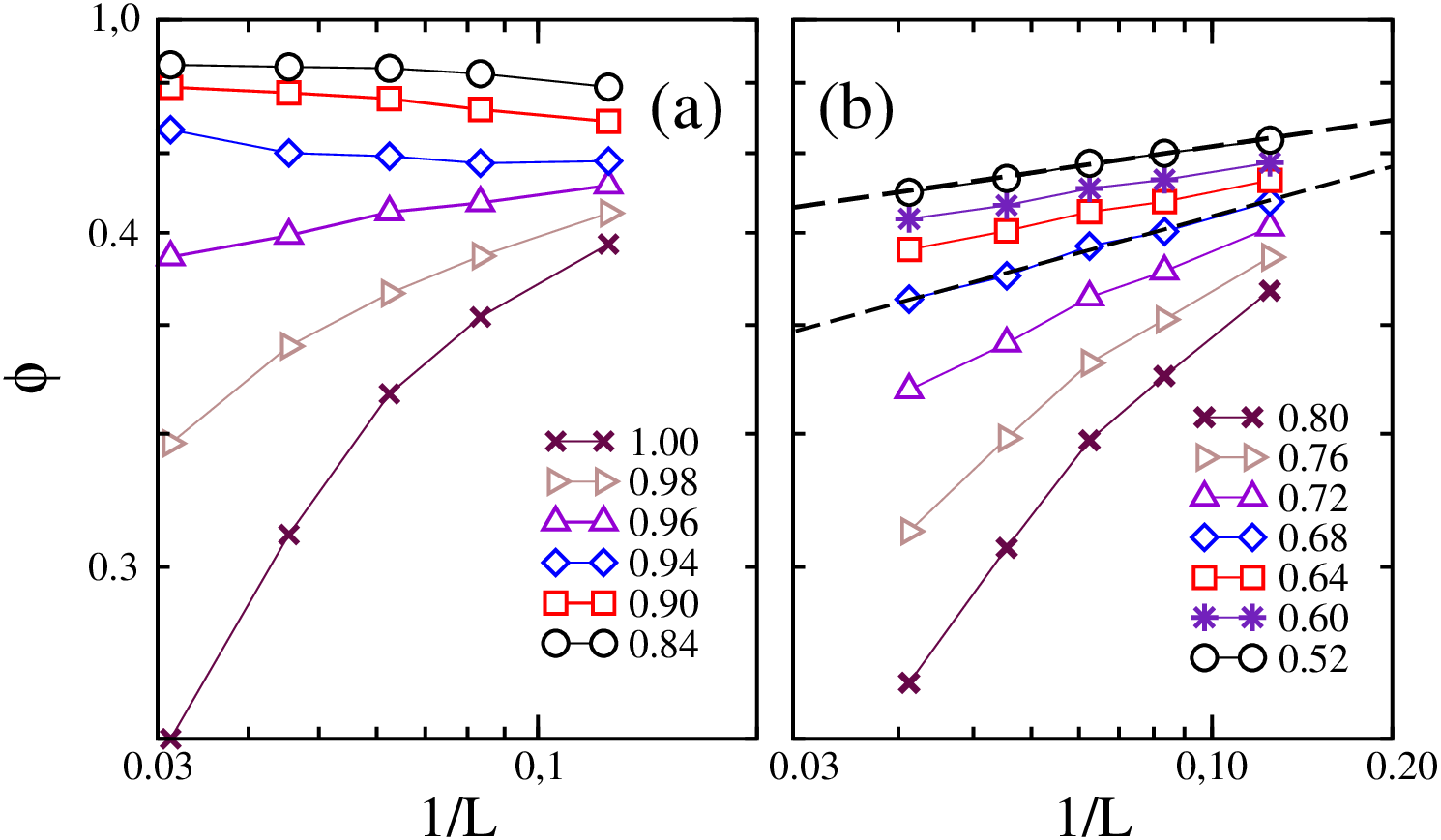}
\caption{(a) Log-log plot of the order parameter
versus $1/L$ for $\Delta=0.4$ and the temperatures
indicated in the figure. Error bars are smaller than the size
of the symbols. (b) Same as in (a) but for
$\Delta=\pi/4$. Dashed thick lines in (b)
stand for linear regression of the data. Thin solid lines are guides
to the eyes.}
\label{fig3}
\end{figure}

In the inset of Fig.~1(a) we plot the specific heat $C$
vs $T$ for $\Delta=0.4$. Clearly a singularity develops
as $L$ increases. We observe a weakening of this singularity as
we increase $\Delta$ from $0$ to $0.7$. For $\Delta \approx \pi/4$ 
we see that the singularity washes out completely (see inset of Fig.~ 1(b)).
In Fig.~ 2 we show a log-log plot of $C_{max}$ versus $T$, where $C_{max}$
is the value of
the specific heat at its maximum. We obtain $\alpha/\nu$ from the
straight line slopes using $C_{max} \propto L^{\alpha/\nu}$.
This is so for
$\Delta =0.2, 0.4$ and $0.6$ for which we obtain 
$\alpha/\nu=0.83(4), 0.63(4)$ and and $0.31(6)$ respectively. 
In contrast, data plotted in Fig.~2 for $\Delta > 0.6$ show not
straight lines  but curves that become flatter as $L$ increases
suggesting that $\alpha/\nu=0$.
This dependence of $\alpha/\nu$ with $\Delta$ is 
in agreement with the Harris criterion \cite{harris}, in 
principle valid only for systems with short ranged interactions.

\begin{figure}[!t]
\includegraphics*[width=85mm]{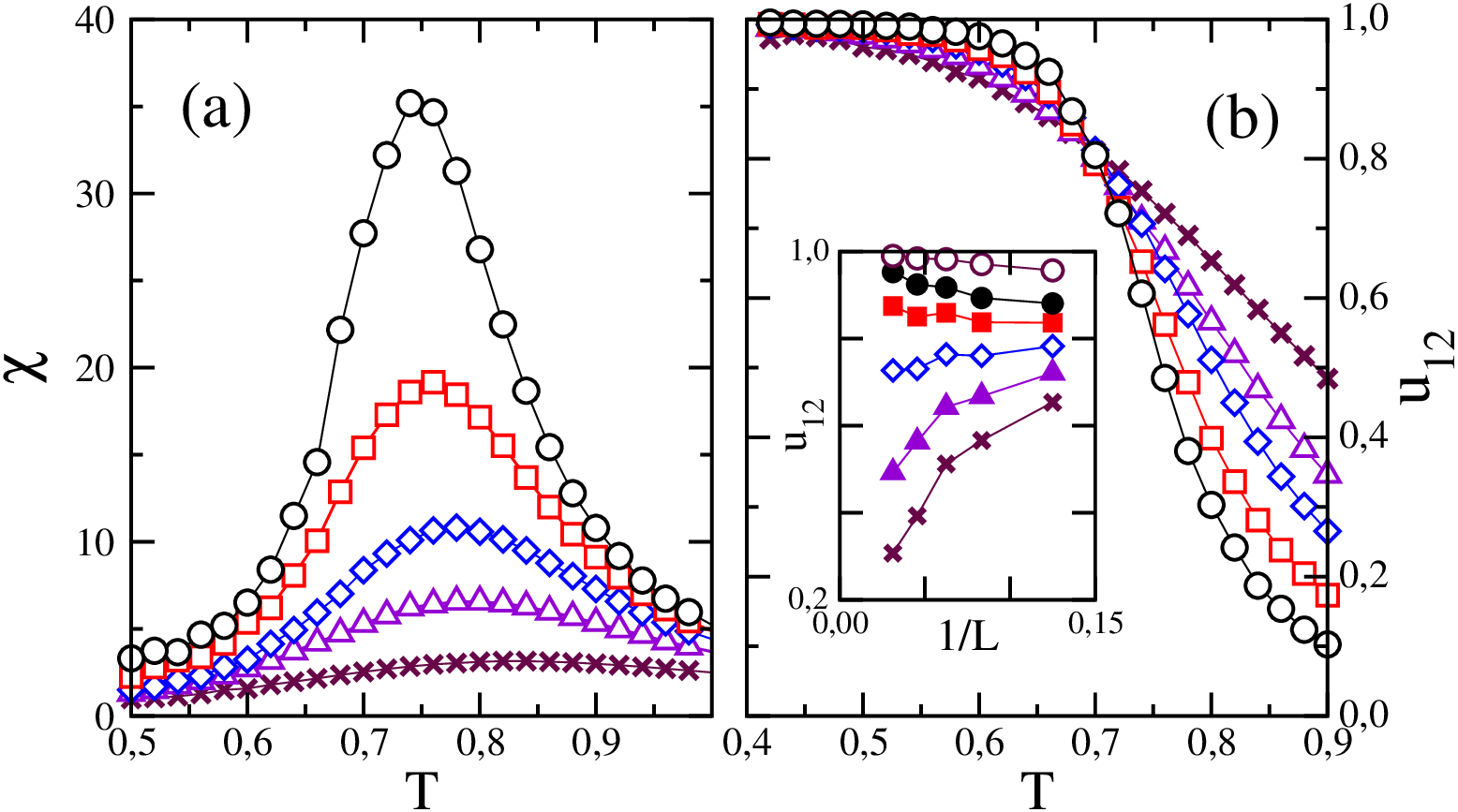}
\caption{(a) Susceptibility versus $T$ for $\Delta=\pi/4$.
$\times$, $\bigtriangleup$, $\diamond$, $\Box$ and
$\circ$  
stand for
$L=8, 12, 26, 22$ and $32$ respectively Lines are only guides to the eyes.
(b) Cumulant $u_{12}$ versus $T$ for $\Delta=\pi/4$. Symbols stand
for the same system sizes as in (a). In the inset
of Fig. (b), $u_{12}$ versus $1/L$ for different temperatures and
$\Delta=\pi/4$. $\circ$, $\bullet$, $\blacksquare$,
$\diamond$,
$\blacktriangle$
and $\times$ stand for $T=0.80, 0.76, 0.72, 0.68, 0.64$
and $0.52$ respectively.
}
\label{fig4}
\end{figure}

We next discuss whether strong or weak long-range order exists at low
temperatures for $\Delta=\pi/4$. 
For this purpose, we compare log-log plots of $\phi$ vs $1/L$
for $\Delta=0.6$ and $\Delta=\pi/4$.
These plots are shown in Figs.~3(a) and 3(b) respectively 
and are qualitalively different. Fig.~3(a) is consistent with a
phase transition from a magnetically disorder state above
$T=0.95(1)$ to a strong long range order below them. 
We have obtained similar behaviour for other values of $\Delta$ in
the  $(0, 0.7)$ range.  On the other hand, in Fig.~3(b) $\phi$ appears to 
decrease algebraically with $L$ over a wide range of temperatures for
$T\le 0.68$
Dashed thick lines in  the figure stand for linear
regression of the data for $T=0.52$ and $0.68$ and their slope 
gives $\beta/\nu= 0.06(2)$ and $0.12(4)$ respectively 
using $\phi \propto L^{\beta/\nu}$. 

We also gathered information from the behaviour of
susceptibility $\chi$. In Fig.~4(a), plots of $\chi$ vs $T$ for $\Delta=\pi/4$
show curves whose peak grows as $L$ increases.
From the data shown, we obtain $\chi _{max} \propto L^{1.85(5)}$ where 
$\chi _{max}$ stands for the maximum value of $\chi$ versus $T$ for a
given value of $L$.
We obtain qualitatively similar results for $\Delta < \pi/4$. Note also
in Fig.~4(a) that the position of the maximum of $\chi$, $T_m$,  
changes with $L$. In the inset of Fig.~5 we plot $T_m$ 
versus $1/L$ for different values of $\Delta$. 
Direct extrapolation of these data to $1/L \to 0$ gives an 
estimation of the transition temperature $T_c$. The resulting values
of $T_c$ vs $\Delta$ are plotted in Fig.~5.

A  more accurate determination of $T_c$ and some additional information 
about the nature of the transition was obtained from 
$u_{12}$. In Fig.~4(a) we plot $u_{12}$ vs $T$ for
$\Delta=\pi/4$. Curves for different values of $L$ exhibit a clear
crossing at $T_c=0.69(1)$ and $u_{12}=0.82(1)$. We have obtained similar
plots for various values of $\Delta$ that enabled us to estimate 
$T_c$ by the value of $T$ where the curves cross. The resulting
values of $T_c$ are plotted vs $\Delta$ in
Fig.~5, which gives the global phase diagram for our model. In the inset of
Fig.~4(b) we plot $u_{12}$ versus $1/L$ for $\Delta=\pi/2$. For the
system sizes we have considered, curves exhibit marked
finite size effects even well below $T = 0.69$. Results for 
$T=0.52$ and $0.64$ shown in this inset seem to indicate 
that $u_{12} \to 1$ as $1/L \to 0$ or at least that
$1 \le <\phi ^2>/<\phi>^2 < 1.02$. This is consistent with the behaviour \cite{koster} 
found for $u_{12}$ for 2D $XY-like$ models that exhibit a Kosterlitz-Thouless 
transition with {\it weak} long-range order below a transition 
temperature.

In sum, we have reported evidence from MC simulations that disorder
in the orientation of the quadrupolar anisotropy axes on systems
of interacting dipoles in square lattice is relevant,
in the sense
that modifies the critical behavior of the thermal transition between a 
paramagnetic and the antiferromagnetic dipolar phase.

\begin{figure}[!t]
\includegraphics*[width=88mm]{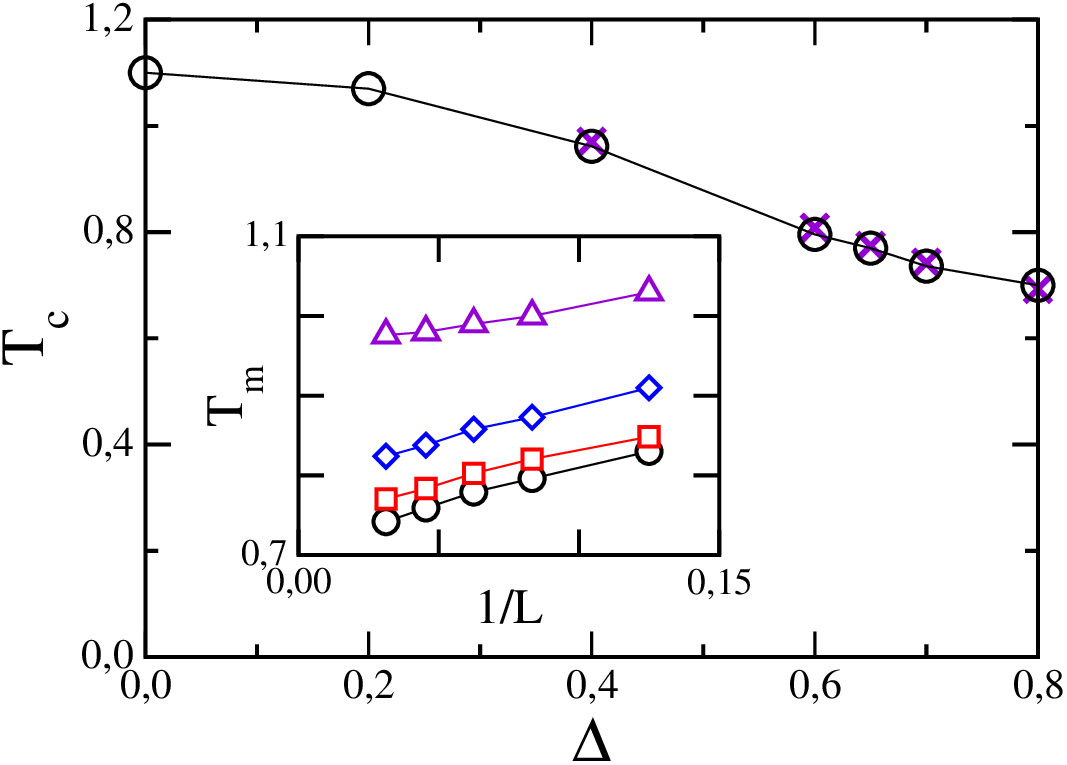}
\caption{Transition temperature $T_c$ versus $\Delta$. $\bigcirc$ 
stand for values of $T_c$ obtained from extrapolation of 
temperatures at which susceptibility attains its maximum for 
different values of $L$ (see text). $\times$ stand for $T_c$ obtained
as the temperature for which $u_{12}$ is independent of $L$.
In the inset, temperature of maximum susceptibility versus
$1/L$. $\triangle$, $\diamond$, $\square$ and $\circ$ stand 
for $\Delta = 0.4, 0.6, 0.7$ and $\pi/2$ respectively.} 
\label{fig1}
\end{figure}

We enjoyed interesting discussions with J. F. Fern\'andez and are
grateful to Institute Carlos I at University of Granada 
for much computer time. We thank financial support from Grant FIS2006-00708
from the Ministerio de Ciencia e Innovaci\'on of Spain.


\end{document}